\documentclass[12pt]{iopart}

\usepackage{color}
\usepackage{graphics}
\usepackage{amsfonts}
\usepackage{epsfig}

\begin{document}

\catcode`\ä = \active \catcode`\ö = \active \catcode`\ü = \active
\catcode`\Ä = \active \catcode`\Ö = \active \catcode`\Ü = \active
\catcode`\ß = \active \catcode`\é = \active \catcode`\è = \active
\catcode`\ë = \active \catcode`\ô = \active \catcode`\ê = \active
\catcode`\ø = \active \catcode`\ò = \active \catcode`\í = \active
\catcode`\Ó = \active \catcode`\ú = \active \catcode`\á = \active
\catcode`\ã = \active \catcode`\à = \active
\defä{\"a} \defö{\"o} \defü{\"u} \defÄ{\"A} \defÖ{\"O} \defÜ{\"U} \defß{\ss} \defé{\'{e}}
\defè{\`{e}} \defë{\"{e}} \defô{\^{o}} \defê{\^{e}} \defø{\o} \defò{\`{o}} \defí{\'{i}}
\defÓ{\'{O}} \defú{\'{u}} \defá{\'{a}} \defã{\~{a}}\defà{\`{a}}

\newcommand{\li}{$^6$Li}
\newcommand{\EFup}{E_{F\uparrow}}
\newcommand{\EFdown}{E_{F\downarrow}}
\newcommand{\TFup}{T_{F\uparrow}}
\newcommand{\TFdown}{T_{F\downarrow}}
\newcommand{\Tup}{T_{\uparrow}}
\newcommand{\Tdown}{T_{\downarrow}}
\newcommand{\nup}{n_{\uparrow}}
\newcommand{\ndown}{n_{\downarrow}}
\newcommand{\normTau}{\tilde{\tau}}
\newcommand{\Zupavg}{\langle Z_\uparrow \rangle}
\newcommand{\Zdownavg}{\langle Z_\downarrow \rangle}
\newcommand{\Zupdownavg}{\langle Z_{\uparrow(\downarrow)} \rangle}
\newcommand{\Zup}{Z_\uparrow}
\newcommand{\Zdown}{Z_\downarrow}
\newcommand{\Zupdown}{Z_{\uparrow(\downarrow)}}
\newcommand{\Veff}{V_{\mathrm{eff}}}
\newcommand{\Vsigeff}{V_{\sigma,\mathrm{eff}}}
\newcommand{\vdrift}{v_{\rm drift}(\vec{r})}
\newcommand{\mitcua}{Department of Physics, MIT-Harvard Center for Ultracold Atoms, and Research Laboratory of Electronics,
        MIT, Cambridge, Massachusetts 02139, USA}
\newcommand{\lens}{LENS and Dipartimento di Fisica, Università di Firenze, and INO-CNR, 50019 Sesto Fiorentino, Italy}

\title{Spin Transport in Polaronic and Superfluid Fermi Gases}
\author{Ariel Sommer, Mark Ku, and Martin W. Zwierlein}
\address{\mitcua}

\begin{abstract}
We present measurements of spin transport in ultracold gases of fermionic \li{} in a mixture of two spin states at a Feshbach resonance. In particular, we study the spin dipole mode, where the two spin components are displaced from each other against a harmonic restoring force. We prepare a highly-imbalanced, or polaronic, spin mixture with a spin dipole excitation and observe strong, unitarity limited damping of the spin dipole mode. In gases with small spin imbalance, below the Pauli limit for superfluidity, we observe strongly damped spin flow despite the presence of a superfluid core.
\end{abstract}
\maketitle
\section{Introduction}
The quality of transport is one of the most important properties distinguishing states of matter. Of great technical importance, electrons in condensed matter materials can flow as currents or supercurrents, or be localized in an insulator, or even switch their state of conductivity through controllable parameters like an applied magnetic field. The task of many-body physics is to develop models that may explain the observed transport properties in a system. Dilute atomic gases cooled to quantum degeneracy provide ideal systems for testing many-body theories. In particular, Feshbach resonances~\cite{chin10fesh} in atomic Fermi gases allow experimental control over the strength of two-body interactions, giving access to the BEC-BCS crossover regime~\cite{kett08varenna,gior08review}. Transport properties have played an important role in characterizing strongly interacting Fermi gases in the BEC-BCS crossover, with the observation of hydrodynamic flow indicating nearly perfect fluidity~\cite{cao10univ, ohar02science}, the measurement of collective excitation frequencies probing the equation of state~\cite{kina04sfluid,bart04coll,altm07precision}, and the observation of vortex lattices in rotating gases demonstrating superfluidity~\cite{zwie05vort}. However, there have been few studies of the transport of spin in strongly-interacting Fermi gases. Spin transport has been used in weakly-interacting systems to probe the onset of quantum degeneracy~\cite{dema02spin} and to generate spin waves~\cite{du08obse}.

Here we study spin transport in strongly interacting two-component Fermi gases. Spin currents are strongly damped in such systems due to the high collision rate between opposite spin atoms: as two-body scattering does not conserve relative momentum, each scattering event on average reduces the net spin current~\cite{dami00theo}. At the Feshbach resonance, scattering is maximal, with a mean free path between collisions of opposite spins that can be as short as one interparticle spacing - the smallest possible in a three-dimensional gas. Recently, we reported measurements of spin transport in strongly interacting Fermi gases with an equal number of atoms in two spin states~\cite{somm11univ}. We demonstrated that interactions can be strong enough to reverse spin currents, with two clouds of opposite spin almost perfectly repelling each other. The spin diffusivity was found to reach a lower limit on the order of $\hbar/m$ at unitarity, the quantum limit of diffusion. Here we consider the case where the number of atoms in the two states is unequal, and study spin transport in the polaron and phase-separated superfluid regimes.
In highly polarized systems that remain non-superfluid down to zero temperature~\cite{zwie05imbalance,shin06phase,schi09polaron}, spin currents are expected to become undamped due to Pauli blocking~\cite{bruu08coll,reca08spinpol,pari09evap}. In this imbalanced regime, a high-frequency mode observed after a compressional excitation was interpreted as a weakly damped spin quadrupole (or breathing) mode~\cite{nasc09imbal}. The question of the damping properties of the spin excitation and its temperature dependence was left open. Spin transport properties of ultracold Fermi gases have been investigated theoretically in~\cite{poli07spin, rain08spin,bruu08coll,reca08spinpol,pari09evap,duin10spin}, allowing direct comparison between theory and experiment.

In section {\bf \ref{sec:polaron}} we present measurements of the damping rate of spin excitations in highly polarized Fermi gases as a function of temperature. We show that damping is maximal at finite temperatures. In section {\bf \ref{sec:sf}} we study smaller spin polarizations, below the Pauli limit of superfluidity~\cite{zwie05imbalance}, just enough to reveal the presence of a superfluid core in the system. We show that the spin dipole mode is strongly damped despite the presence of the superfluid. At first this is surprising, as one might expect the superfluid to flow without friction, and we discuss an explanation for this observation.

\section{\label{sec:polaron}Highly Imbalanced Fermi Gases}
Fermi gases with resonant interactions can remain normal down to zero temperature if the spin imbalance exceeds the Pauli (or Clogston-Chandrasekhar) limit~\cite{clog62,chan62,zwie05imbalance,shin06phase,nasc09imbal}. We refer to the spin state with the larger population of atoms as the majority, or spin up state, and the state with fewer atoms as the minority, or spin down state. Radio-frequency (RF) spectroscopy~\cite{schi09polaron} on such systems confirms the quasi-particle picture~\cite{chevy06universal,comb07polaron,prok08polaron} where minority atoms are dressed by the majority Fermi sea, forming a quasi-particle known as the Fermi polaron. The energy of a single polaron in a zero-temperature Fermi sea of spin up atoms has been described using the effective Hamiltonian~\cite{lobo06,bruu08coll,reca08spinpol}
\begin{equation}
\label{eqn:polaronH0}
H = -\alpha \mu_\uparrow + \frac{\mathbf{p}^2}{2m^*},
\end{equation}
where $\mathbf{p}$ is the momentum of the polaron, $m^*$ is the polaron effective mass, $\mu_\uparrow$ is the local spin up chemical potential, and $\alpha$ characterizes the polaron binding energy.
The parameters $\alpha$ and $m^*/m$, where $m$ is the bare mass of spin up and spin down fermions, have been measured experimentally ~\cite{shin08dete,schi09polaron,nasc09imbal,nasc10thermo} and calculated theoretically~\cite{lobo06,comb08fullpolaron,prok08boldpolaron,pila08phase}, giving $\alpha= 0.62$ and $m^*/m\approx$ 1.2 at zero temperature.

We consider a mixture of $N_\uparrow$ spin up fermions and $N_\downarrow$ spin down fermions at temperature $T$ with equal masses and resonant interactions in a spin-independent potential of the form
\begin{equation}
\label{eqn:V}
V(r,z) = \frac{1}{2} m\omega_z^2 z^2 + V_r(r),
\end{equation}
where $r^2 = x^2 + y^2$. The minority cloud is initially displaced by a small amount $Z_\downarrow(0)$ along the $z$ axis and is allowed to relax to its equilibrium position.

In the limit $N_\downarrow \ll N_\uparrow$, the motion of the spin up cloud due to momentum absorbed from the spin down cloud may be neglected. The equation of motion of the spin down center of mass $Z_\downarrow$ is then~\cite{bruu08coll}
\begin{equation}
\label{eqn:poleom}
    m^* \ddot{Z}_\downarrow + (1+\alpha)m\omega_z^2 Z_\downarrow + m^*\dot{Z}_\downarrow/ \tau_P =0,
\end{equation}
where the factor of $(1+\alpha)$ is due to the attraction of the minority fermions to the majority cloud, and $1/\tau_P$ is the momentum relaxation rate due to collisions and is equivalent to the spin drag coefficient~\cite{dami00theo,poli07spin}. By dimensional analysis, $\hbar/\tau_P$ must be given by the majority Fermi energy times a universal dimensionless function of the reduced temperature $T/\TFup$ and the ratio $\TFdown/\TFup$ of the Fermi temperatures, where $T_{F\uparrow(\downarrow)}$ is the majority (minority) Fermi temperature.
The first two terms in (\ref{eqn:poleom}) follow from (\ref{eqn:polaronH0}), (\ref{eqn:V}), and the local density approximation, while the third term is due to damping and is not captured in (\ref{eqn:polaronH0}). In Eq. (\ref{eqn:poleom}) it is assumed that the minority cloud is small enough in size compared to the majority cloud that variation in $1/\tau_P$ with density may be neglected. Also, (\ref{eqn:poleom}) neglects a possible back-action of the minority on the majority atoms that might deform the majority density profile.

In our experimental realization of this transport problem we use a gas of ultracold fermionic \li{} atoms. The \li{} atoms are cooled sympathetically with $^{23}$Na~\cite{hadz03big_fermi} and loaded into a hybrid optical and magnetic trap with an adjustable bias magnetic field~\cite{zwie06direct}. The magnetic field curvature provides essentially perfect harmonic confinement along the axial ($z$) direction, while the optical dipole trap (laser wavelength 1064 nm, waist 115 $\mu$m) provides trapping in the radial directions, with negligible contribution to the axial confinement. With this system we perform a collection of time series measurements. In each time series we prepare the system in a chosen initial state and observe its evolution.

At the Feshbach resonance at 834 G, the magnetic moments of ``spin up" and ``spin down" atoms, the two lowest hyperfine states of \li{}, are equal to 1 part in 1000, as their electron spin is in fact aligned with the magnetic field. Inducing a spin current is therefore extremely challenging on resonance. However, at lower fields, their magnetic moments differ, allowing separation of the two gas clouds by a magnetic field gradient. Our experimental procedure for producing these separated clouds is as follows.

We prepare the system starting with about $1\times 10^7$ atoms of \li{} in the lowest hyperfine state, at a total magnetic field of 300 G. A small fraction of atoms are transferred to the second lowest hyperfine state using a RF Landau-Zener sweep. The mixture is then evaporatively cooled for a variable amount of time by lowering the depth of the optical dipole trap from $k_B  \times  7 \mu$K to a variable final depth between $k_B  \times  0.5 \mu$K and $k_B  \times 1 \mu$K, where $k_B$ is the Boltzmann constant. The optical dipole trap depth is then raised to $k_B \times 6 \mu$K, where the zero temperature Fermi energy in the majority state is between $k_B  \times  0.8 \mu$K and $k_B  \times  1.3 \mu$K.

After the spin mixture is prepared at 300 G, the total magnetic field is reduced gradually over 500 ms to 50 G, where the ratio of the magnetic moments of the two states is 2.5 and interactions are very weak. A magnetic field gradient is applied along the $z$ direction for about 4 ms, imparting a linear momentum of the same sign but a different magnitude to each spin state. The clouds are then allowed to evolve for about 30 ms, and they execute about half of an oscillation period at different amplitudes and frequencies (the frequency ratio is 1.6 between spin up and down). When the clouds have returned to the center of the trap, their centers of mass are displaced from each other by about 200 $\mu$m (for comparison, the $1/\mathrm{e}$ radius of the majority cloud in the $z$ direction is between 200 and 300 $\mu$m at this point). A second gradient pulse is applied along the same direction to remove the relative velocity of the two clouds. The second pulse also removes most of the total center of mass motion. The total magnetic field is then ramped to the Feshbach resonance at 834 G in about 5 ms. At resonance, the two spin states have identical trapping frequencies of 22.8 Hz~\footnote{The system as a whole oscillates harmonically along the $z$ direction at 22.8 Hz due to the residual center of mass energy. This motion does not affect the dynamics in the total center of mass frame because the trapping potential is harmonic in the $z$ direction, and therefore, according to Kohn's theorem, the dynamics in the total center of mass frame are equivalent to the dynamics of a system at rest~\cite{dobs94kohn}.}.

To reach low temperatures, we apply a variable amount of evaporative cooling by lowering the depth of the optical dipole trap after reaching 834 G. The time available for evaporative cooling is limited to about 0.4 s by the relaxation time of the spin excitation. To reach high temperatures, we prepare a hotter cloud at 300 G and heat the system further at 834 G by releasing the atoms from the optical dipole trap and recapturing them. The depth of the optical dipole trap is then ramped gradually to a final value in 80 ms. The final depth is chosen to keep the number of atoms and the temperature approximately constant during the subsequent evolution, and corresponds to an effective radial trap frequency ranging from 80 Hz for the low temperature data to 250 Hz for the high temperature data. After preparing the system at the chosen temperature and with a non-zero spin dipole moment, we are left with typically $N_\uparrow \approx 4\times10^5$ atoms in the majority state and $N_\downarrow \approx 4\times10^4$ atoms in the minority state. We then allow the system to evolve for a variable wait time $t$ before measuring the densities of the spin up and spin down clouds using resonant absorption imaging. Note that we limit the population of the majority cloud to ensure that the central optical density is less than 2, allowing for accurate density measurements.

\begin{figure}
\includegraphics{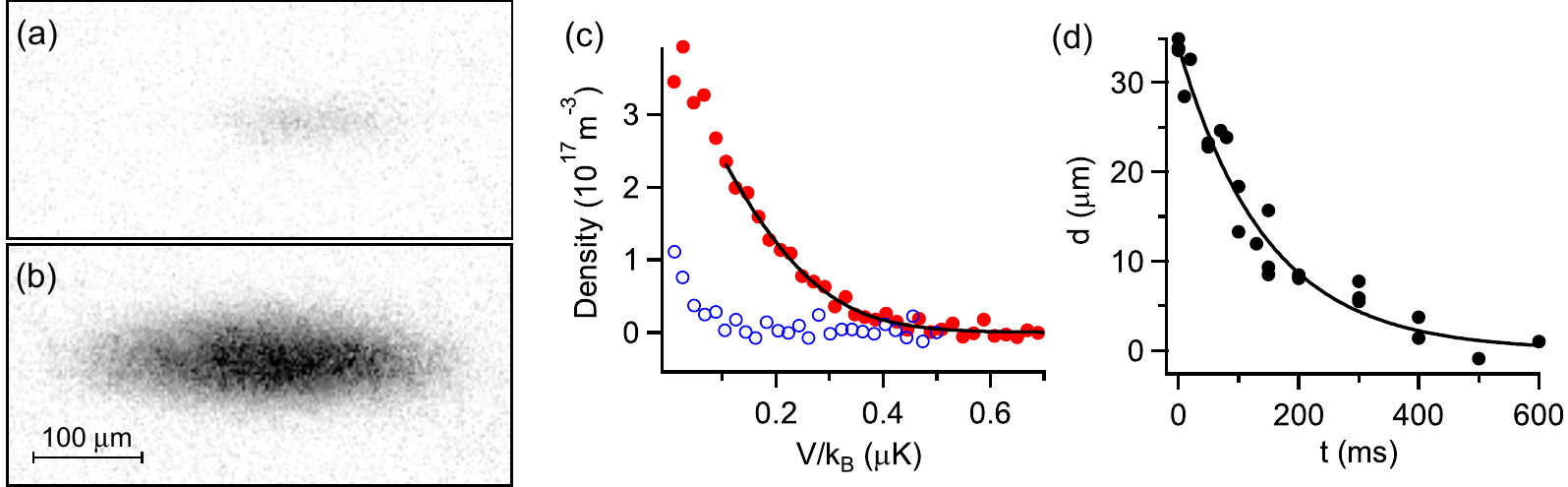}
\caption{Measuring the spin dipole mode of a highly-polarized Fermi gas. {\bf (a)} and {\bf (b)} show two-dimensional column density images of the minority and majority spin state, respectively, obtained using resonant absorption imaging in one run of the experiment. The imaging pulses are each 4 $\mu$s in duration and separated by 6 $\mu$s. The distance between the centers of mass in {\bf (a)} and {\bf (b)} is 34 $\mu$m. {\bf (c)} Density of the majority (red circles) and minority (open blue circles) versus potential energy for $z<\Zup$, obtained from the images in {\bf (a)} and {\bf (b)}. The black line shows a fit to the majority density at the edge of the cloud where the minority fraction is 5\% or less. {\bf (d)} Displacement $d$ of the minority center of mass relative to the majority center of mass as a function of time. This time series includes the run displayed in {\bf (a-c)}. The curve shows an exponential fit.}
\label{fig:polmeth}
\end{figure}
Figs.~\ref{fig:polmeth}{\bf a},{\bf b} show typical two-dimensional column densities of the two spin states after evaporative cooling on resonance. From the column densities we reconstruct the three-dimensional densities $n_\sigma(r,z)$ of each state $\sigma=\uparrow,\downarrow$ using the inverse Abel transformation. The temperature of the system is determined by fitting the majority density as a function of potential energy to the equation of state of a non-interacting Fermi gas~\cite{zwie06direct,shin07phasediagram} (Fig.~\ref{fig:polmeth}{\bf c}): $n_{\uparrow, FG} = -\lambda^{-3} \zeta_{3/2}\left(-e^{\beta(\mu-\Veff)}\right)$, where $\lambda=\sqrt{2\pi\hbar^2/m k_B T}$ is the thermal de Broglie wavelength, $\beta=1/k_B T$, the fit parameters are the chemical potential $\mu$ and the temperature $T$, $\zeta_{3/2}$ is the polylogarithm of order 3/2, and $\Veff =V(r,z-\Zup)$ is the effective potential energy. The fit is restricted to the outer edges of the majority cloud, where $n_\downarrow / n_\uparrow < x_c$. We used a cut-off minority fraction of $x_c = 0.05$ for all clouds with $T < 0.5\TFup$. For some of the data with $0.5<T/\TFup<1$, $x_c$ was increased to 0.08 to increase the available signal, while for the data with $T > 2 \TFup$, $x_c$ was increased to 0.15 for the same reason. These increases in $x_c$ should not affect the accuracy of the thermometry because the system interacts less strongly at high $T/\TFup$. This is demonstrated by our spin susceptibility measurements for the balanced case in~\cite{somm11univ} that agree with the compressibility above $T/T_F \approx 1$, showing the absence of spin correlations in this temperature regime. Also, measurements on the density profiles of imbalanced clouds~\cite{shin08,shin08dete,nasc10thermo} indicate that the majority density is not affected by the presence of minority atoms at high reduced temperatures. For normalization the central densities $n_\sigma(0)$ of each species are recorded and used to define the central Fermi energies $E_{F\sigma} = \hbar^2 k_{F\sigma}^2/{2 m_\sigma}$, with $k_{F\sigma} = (6\pi^2 n_\sigma(0))^{1/3}$, $m_\uparrow=m$, and $m_\downarrow=m^*$, and Fermi temperatures $T_{F\sigma}=E_{F\sigma}/k_B$.

Spin transport is measured by observing the time evolution of the center of mass separation $d(t)=\Zdown(t)-\Zup(t)$ (Fig.~\ref{fig:polmeth}{\bf d}), with $\Zupdown(t)$ the center of mass of the majority (minority) cloud along the $z$ axis at time $t$, determined from a two-dimensional gaussian fit to the column density. We find that $d$ relaxes exponentially to zero, corresponding to an overdamped spin dipole mode, and fit the evolution to an exponential function $d(t) = d_0 \mathrm{e}^{-t/\tau}$. The fitted spin dipole relaxation times $\tau$ can be usefully normalized using $\omega_z$ and $\EFup$ by referring to equation (\ref{eqn:poleom}). When $(\omega_z\tau)^2\gg 1$, the first term in (\ref{eqn:poleom}) may be neglected. In our measurements, $\tau$ is at least 100 ms, so $(\omega_z\tau)^2 > 200$. The product $\omega_z^2\tau$ is then related to fundamental properties of the gas via
\begin{equation}
    \omega_z^2\tau = \frac{m^*/m}{(1+\alpha)\tau_P}.
    \label{eqn:tau}
\end{equation}
The quantity $\normTau=\hbar \omega_z^2\tau/\EFup$ then provides a dimensionless measure of the relaxation time in our experiment, and can be compared to theoretical calculations.

\begin{figure}
    \begin{centering}
    \includegraphics{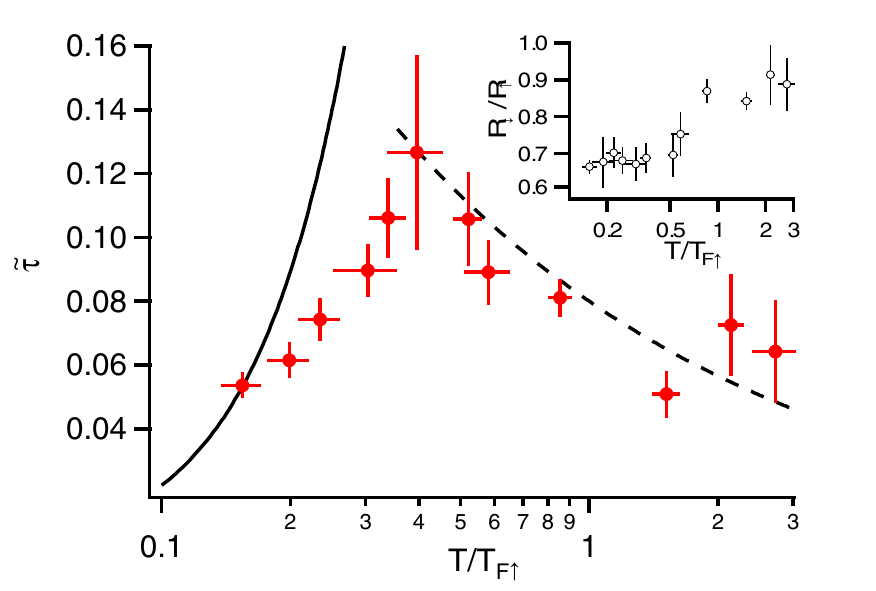}
    \caption{Normalized relaxation time of the spin dipole mode of a highly-polarized Fermi gas as a function of the reduced temperature $T/\TFup$. $\TFup$ is the local Fermi temperature at the center of the majority cloud. The solid curve is the low temperature limit from~\cite{bruu08coll}, given by equation (\ref{eqn:predict}). The dashed curve is the expression $0.08\sqrt{\frac{\TFup}{T}}$.
    The inset shows the average ratio of the minority cloud size to the majority clouds size as a function of the reduced temperature $T/\TFup$. The cloud sizes are defined as the $1/\mathrm{e}$ radii along the $z$ axis, estimated by fitting a two-dimensional gaussian function to the column densities of the two spin states.
    In both figures, each point is a weighted average of the results from 1 to 3 time series, with each time series containing on average 30 spin up-spin down image pairs. The error bars give standard deviations due to statistical fluctuations within a time series. Where the results of more than one time series are averaged, the error bars show the standard deviation of the weighted mean, determined from the standard deviations from each time series.
    }
    \label{fig:polaronSD}
    \end{centering}
\end{figure}
Figure \ref{fig:polaronSD} shows the measured values of the dimensionless relaxation time $\normTau$ as a function of the reduced temperature $T/\TFup$. $\normTau$ increases at low temperatures before reaching a maximum of 0.13(3)$\EFup$ for $T/\TFup = 0.40(6)$, and decreases at higher temperatures.  We interpret the behavior of the relaxation time at low temperatures as a consequence of Pauli blocking: as the temperature is lowered significantly below the majority Fermi temperature, the phase space available for a minority atom to scatter goes to zero. The reduction in $\normTau$ at high temperatures is expected: at high temperatures, $1/\tau_P$ is essentially given by the collision rate in the gas~\cite{vich99coll}, $1/\tau_P \sim n \sigma v$. The scattering cross section $\sigma$ on resonance for $T\gg \TFup$ is given by the square of the de Broglie wavelength and is thus proportional to $1/T$, while the average speed $v$ of the particles is proportional to $\sqrt{T}$. Hence $\normTau$ is expected to decrease like $\hbar n\sigma v/\EFup \propto \sqrt{\TFup/T}$. We observed behavior similar to Fig.~\ref{fig:polaronSD} in three-dimensional Fermi gases with resonant interactions and equal spin populations in~\cite{somm11univ}, although we see more significant Pauli blocking here than in~\cite{somm11univ} at comparable temperatures.

It would be interesting to have available a calculation of spin transport coefficients such as $1/\tau_P$ at arbitrary temperatures for comparison with our data. A full solution is available for Fermi gases with equal populations in one spatial dimension~\cite{poli07spin, rain08spin} and shows qualitative features similar to our data, showing a maximum of the spin drag coefficient (analogous to $\normTau$) at finite temperatures on the order of $T_F$.

We expect our data to differ quantitatively from predictions for a homogeneous system. The measured quantity $\normTau$ is a global property of the trapped system, while the momentum relaxation rate $1/\tau_P$ is a local quantity. For $T\gg\TFup$, $1/\tau_P\propto n_\uparrow$, and $1/\tau_P$ increases with increasing majority density, while for $T\ll\TFup$, due to Pauli blocking $1/\tau_P\propto \EFup (T/\TFup)^2 \propto n_\uparrow^{-2/3}$, and $1/\tau_P$ decreases with increasing majority density. Additionally, the variation of $1/\tau_P$ should cause the spin current to be non-uniform. The effect of inhomogeneity should be greater at high reduced temperatures, where the minority cloud size approaches the majority cloud size. The inset in Fig.~\ref{fig:polaronSD} shows the ratio of the cloud sizes $R_\downarrow/R_\uparrow$ as a function of the reduced temperature, where $R_{\uparrow(\downarrow)}$ is the $1/\mathrm{e}$ width in the $z$ direction from a two-dimensional gaussian fit to the majority (minority) column density. Indeed $R_\downarrow/R_\uparrow$ increases with increasing $T/\TFup$. Even at the lowest temperatures, $R_\downarrow/R_\uparrow$ remains significant, attaining a value of 0.7, due to the finite minority fraction $N_\downarrow/N_\uparrow\approx0.1$. The effect of inhomogeneity is therefore reduced at low temperatures, but should remain present.

We compare our results at low temperatures to the low temperature limit in Ref.~\cite{bruu08coll}, which can be written as
\begin{equation}
    \normTau = {\rm c}\frac{\alpha^2}{1+\alpha}\left(\frac{m^*}{m}\right)^2\left(\frac{T}{\TFup}\right)^2
    \label{eqn:predict}
\end{equation}
for temperatures $T\ll \TFup$~\footnote{We omit a term due to the relative velocity of the spin up and spin down clouds, which produces a correction of less than 1\% in the overdamped, finite temperature regime accessed in this experiment.}. The prefactor ${\rm c}$ changes slightly from ${\rm c} = \frac{2\pi^3}{9} = 6.89\dots$ to ${\rm c} \approx 6.0$ as the temperature rises from far below $\TFdown$, where even the minority cloud is degenerate, to temperatures where $\TFdown\ll T\ll \TFup$ and the minority is a classical gas~\cite{bruu08coll}. In our coldest data, $T\approx 0.5 \TFdown$ and $\TFdown\approx 0.3\TFup$, assuming $m^*=1.2\,m$. To compare our data to Ref.~\cite{bruu08coll} using (\ref{eqn:predict}) we set ${\rm c}= \frac{2\pi^3}{9}$, $\alpha=0.6$ and $m^* = 1.2\, m$. The experimental data agree with the theory at the lowest temperatures measured (see Fig.~\ref{fig:polaronSD}). The deviation at higher temperatures is expected as the $T\ll \TFup$ limit becomes inapplicable. The convergence of the experimental data to the theoretical value at low temperature despite the inhomogeneity of the system may be partly due to the reduced minority cloud size at low temperatures, which reduces the effects of inhomogeneity, as discussed above. Additionally, the variation of the momentum relaxation rate with density will to some extent cancel at moderately low temperatures, as $1/\tau_P$ changes from increasing with increasing density at high reduced temperatures to decreasing with increasing density due to Pauli blocking at low reduced temperatures. The crossing of the experimental curve with the predictions for a uniform system at low temperatures therefore does not necessarily indicate that the inhomogeneity is negligible at low temperatures in this measurement.

At high temperatures $T\gg T_{F\uparrow,\downarrow}$, the spin transport properties of a trapped system can be calculated from the Boltzmann transport equation. For vanishing minority fraction, we find (now with $\alpha=0$ and $m^*=m$ and assuming harmonic trapping in all three directions)~\cite{somm11univ}
\begin{equation}
\normTau = \frac{8}{9\pi^{3/2}\,\epsilon}\sqrt{\frac{\TFup}{T}} \approx \frac{0.16}{\epsilon} \sqrt{\frac{\TFup}{T}},
\label{eqn:boltzmann}
\end{equation}
where $\epsilon=1$ when the drift velocity distribution is uniform. This result features the expected dependence $\propto \sqrt{\frac{\TFup}{T}}$ on temperature. The relative velocity between the two spin states cannot be truly constant in space but has to be depressed in the center, where the density is highest and momentum relaxation is fastest. In general,
\begin{equation}
   \epsilon = \frac{\int {\rm d}^3r \, v_{\rm drift}(\frac{\vec{r}}{\sqrt{2}}) e^{-\beta V}}{\int {\rm d}^3r \,v_{\rm drift}(\vec{r}) e^{-\beta V}},
    \label{eqn:velocityfactor}
\end{equation}
where $V$ is the trapping potential (here assumed to be quadratic), and $\vdrift$ is the average velocity along $z$ of minority atoms at position $\vec{r}$. For example, for a quadratic drift velocity profile, $v_{\rm drift}(\vec{r}) = a x^2 + b y^2 + c z^2$, the predicted $\normTau$ is reduced by factor of $\epsilon=2$. We find that the high temperature result (\ref{eqn:boltzmann}) with $\epsilon= 2$ leads to close agreement with our experimental results (Fig.~\ref{fig:polaronSD}). This model is interesting because it estimates the effects of inhomogeneous density and velocity distributions, but it has shortcomings. The drift velocity should remain non-zero everywhere, rather than going to zero at the origin as in the quadratic case, and should have a radial component. A full quantitative description of the overdamped spin dipole motion in the high temperature limit in an external trapping potential will therefore be more complex.

\section{\label{sec:sf}A Superfluid with Small Spin Polarization}
We extend the method of the previous section to study spin transport in Fermi gases with resonant interactions and small spin imbalance. When the global polarization $\frac{N_\uparrow-N_\downarrow}{N_\uparrow+N_\downarrow}$ is less than about 75\% in a harmonically trapped Fermi gas at low temperature and with resonant interactions, the system phase separates into a superfluid core surrounded by a polarized normal state region~\cite{zwie05imbalance,shin06phase,nasc09imbal}. The superfluid core is visible as a sharp reduction in the density difference of the two spin states~\cite{shin06phase}.
The transition between the superfluid and the imbalanced normal regions forms a sharp interface below a tricritical point, where the density imbalance jumps between the two regions~\cite{shin08}.
Scattering and spin transport at the interface between a normal and superfluid Fermi gas have been considered theoretically in Refs.~\cite{scha07norm,pari09evap}.

\begin{figure}
\includegraphics{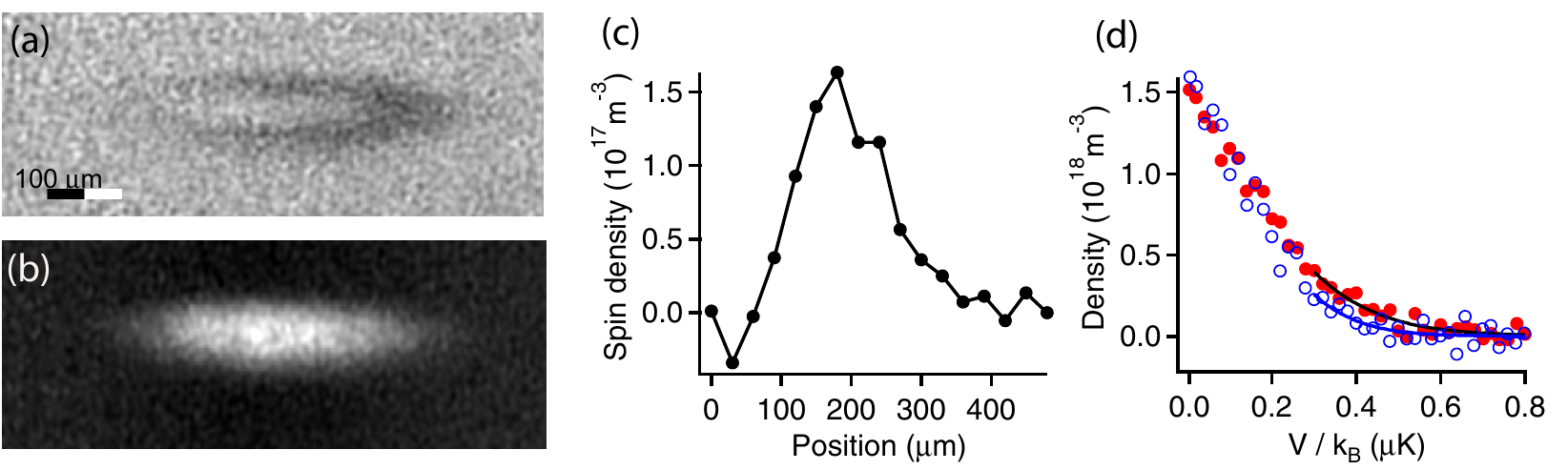}
\caption{Spin dipole mode of an imbalanced Fermi gas with a superfluid core. Phase contrast images are taken with imaging light detuned {\bf (a)} halfway between the resonance frequencies of the two states and {\bf (b)} at large red detuning from both states. The image in {\bf (a)} is proportional to the difference in column densities of the two states. The depletion of the density difference in the center of the cloud indicates the superfluid region. It is displaced from the center of the majority due to the spin dipole excitation.{\bf (c)} shows the difference in reconstructed three-dimensional densities of the spin up and spin down clouds as a function of the $z$ coordinate for $z>0$. The depletion in the center again indicates pairing and superfluidity~\cite{shin06phase}. An elliptical average over a narrow range of the radial coordinate $r$ is used to increase the signal to noise ratio. {\bf (d)} Three-dimensional densities of the two states as a function of the effective potential defined in the text. Solid red circles: majority density, open blue circles: minority density. The curves are fits to the densities using the equation of state of a unitary Fermi gas at zero imbalance to get upper and lower bounds on the temperature.}
\label{fig:sf}
\end{figure}

To observe spin transport in an imbalanced gas containing a superfluid, we prepare a spin mixture with a global polarization of 17(3)\%. The gas is cooled at 300 G and again at 834 G after creating the spin dipole excitation as described in the previous section. Two off-resonant phase contrast images are taken to measure the densities of each spin state. An imaging pulse tuned halfway between the resonance frequencies of the two states directly measures the difference in the column densities (Fig.~\ref{fig:sf}{\bf a}) while a second pulse, red-detuned from both states (Fig.~\ref{fig:sf}{\bf b}), provides additional information needed to reconstruct the total column density in each state~\cite{shin08}. From the column densities of each state we obtain three-dimensional density distributions using the inverse Abel transformation.

The two-dimensional spin density (Fig.~\ref{fig:sf}{\bf a}) and three-dimensional spin density (Fig.~\ref{fig:sf}{\bf c}) show a reduction near the center of the trap, with the three-dimensional density going to zero, characteristic of the superfluid core in imbalanced Fermi gases~\cite{shin06phase}. We have checked that the shell structure remains even after the spin density reaches equilibrium. Additionally, we estimate the temperature $T$ of the system to confirm that it is cold enough to contain a superfluid. In unpolarized systems, the superfluid transition is predicted to occur at about $T_c/T_F = 0.173(6)$~\cite{goul10tc}, where $k_B T_F = E_F =\hbar^2 (6\pi^2 n(0))^{2/3}/2m$ and $n$ is the density per spin state. This theoretical value agrees well with the determination of $T_c/T_F$ by our group~\footnote{Mark Ku, Ariel Sommer, Andr\'e Schirotzek, and Martin W. Zwierlein, in preparation.}. Fitting the equation of state of a unitary Fermi gas at zero imbalance~\cite{ku10equa} to the majority (minority) density gives an estimate $T_{\uparrow(\downarrow)}$ of the temperature. The fits are restricted to $\Vsigeff> 0.3$ $\mu K$, where $\Vsigeff=V(r,z-Z_\sigma)$, to exclude the putative superfluid region. Compared to a balanced gas at unitarity with $N_\uparrow'=N_\downarrow'=N_\uparrow$, and at the same temperature T, the majority cloud should have a larger size because the interaction energy between the spin up and spin down atoms is attractive. We therefore expect that $\Tup$ is an overestimate of $T$. Likewise, we expect $\Tdown < T$, and we consider $\Tup$ and $\Tdown$ to provide approximate upper and lower bounds on $T$.

\begin{figure}
\begin{centering}
\includegraphics{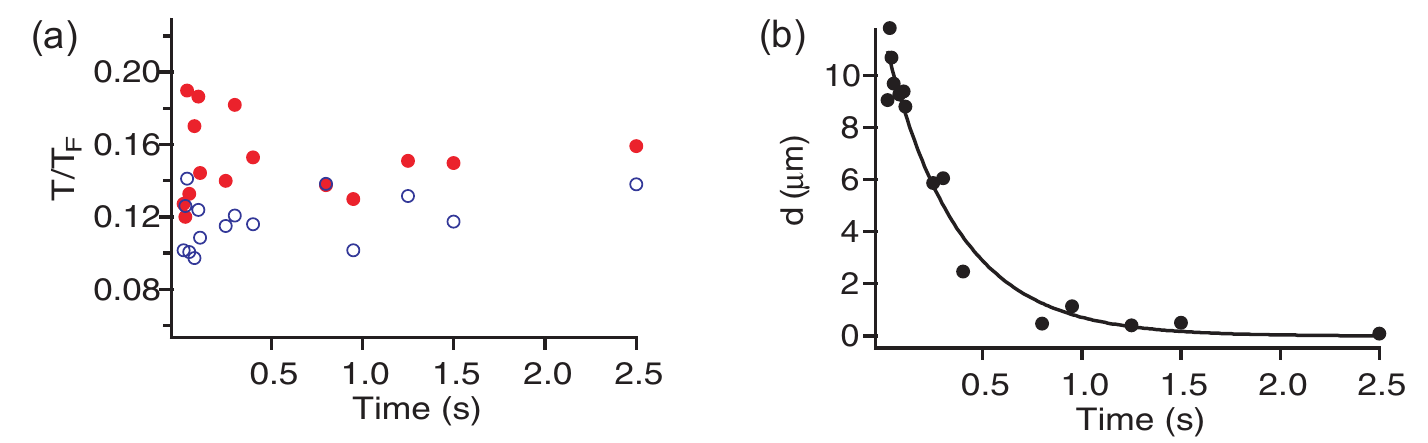}
\caption{{\bf (a)} Reduced temperature as a function of time during relaxation of the spin dipole excitation in a spin polarized Fermi gas containing a superfluid region. Red solid (blue open) circles: $T_{\uparrow(\downarrow)}/T_{F\uparrow(\downarrow)}$ from a fit to the edge of the majority (minority) spin state using the equation of state of an unpolarized unitary Fermi gas, giving an upper (lower) limit to true temperature. {\bf (b)} The displacement of the spin up and spin down centers of mass relaxes exponentially, indicating strong spin drag despite the presence of a superfluid.}
\label{fig:sftime}
\end{centering}
\end{figure}

Figure \ref{fig:sftime}{\bf a} shows the temperature bounds during the approach to equilibrium. Time-averaging gives $0.12(2) < T/T_F < 0.15(2)$, where $T_F \equiv \TFup \approx \TFdown$. The error estimates refer to standard deviations. These bounds confirm that the system is in the vicinity of the superfluid transition. In contrast to the dissipationless flow that defines the superfluid state, however, we observe strong damping of the spin dipole mode. Figure \ref{fig:sftime}{\bf b} shows that the displacement $d$ between the majority and minority centers of mass along the $z$ axis relaxes gradually to zero, rather than oscillating as would be expected in a dissipationless system. The $1/\mathrm{e}$ relaxation time $\tau = 360$ ms corresponds to a spin drag coefficient~\cite{dami00theo, poli07spin} of $\omega_z^2\tau = 0.06(1) \EFup/\hbar$, close to the maximum spin drag coefficient in non-polarized trapped Fermi gases at unitarity~\cite{somm11univ}.

The strong damping of the spin dipole mode, or spin drag, is surprising when one considers that the superfluid moves through a background of spin up atoms at a velocity $\sim d/\tau = 30$ $\mu$m/s, much less than the critical velocity of about $0.3 v_F$~\cite{mill07critical,comb06collective}, where $v_F = \sqrt{2 E_F / m} = 50$ mm/s. However, unlike in a superconductor, where the electrons move through a stationary background of scatterers, here the background is a gas of atoms with a root-mean-square velocity of at least $\sqrt{\frac{3}{5}}v_F$, significantly exceeding the critical velocity. In the inhomogeneous case realized here there is an additional source of damping due to the superfluid-to-normal interface. One then expects that a fraction of atoms in the normal region are reflected at the interface due to the pairing gap~\cite{scha07norm,pari09evap}. The relative importance of the two effects depends on whether a spin current flows through the superfluid or around it. However, we are not able to determine the spatial distribution of the spin current with our present data.

\section{Conclusions}
In this work we presented our measurements on the damping of the spin dipole mode in a highly polarized Fermi gas with resonant interactions, over a wide range of temperatures. The damping is seen to become weaker at temperatures significantly less than the majority Fermi energy, as expected from Pauli blocking, i.e. the fact that quasi-particles in a Fermi liquid become long lived at sufficiently low temperatures. These measurements provide the first quantitative test of theoretical calculations of the spin transport properties of highly polarized Fermi gases. We also observe spin transport in a Fermi gas with low spin polarization containing a superfluid region. It is found that the spin dipole motion remains strongly damped, revealing the importance of friction between the superfluid and the normal component, possibly accompanied by reflection processes at the interface.

This work was supported by the NSF, AFOSR-MURI, AFOSR-YIP, ARO-MURI, ONR, DARPA YFA, a grant from the Army Research Office with funding from the DARPA OLE program, the David and Lucille Packard Foundation and the Alfred P. Sloan Foundation. The authors would like to thank Andr\'e Schirotzek, Giacomo Roati and Peyman Ahmadi for experimental assistance, and David Huse for interesting discussions.
\\
\bibliography{sommer_spin}
\end{document}